\documentclass[preprint,10pt]{elsarticle}
\usepackage{epsfig}
\usepackage{graphicx}
\usepackage{amsmath}
\usepackage{amssymb}
\usepackage{multirow}
\usepackage{subfigure}
\usepackage[breaklinks=true,bookmarks=false]{hyperref}
\usepackage{enumerate,microtype}
\usepackage{color}
\usepackage{longtable}

\newcommand{\ignore}[1]{}
\journal{Journal Name}

\begin{document}

\begin{frontmatter}

\title{Single Image Super Resolution based on a Modified U-net\\ with Mixed Gradient Loss}

\author{Zhengyang Lu}
\author{Ying Chen}

\address{Key Laboratory of Advanced Process Control for Light Industry (Ministry of Education),\\ Jiangnan University}
\address{Wuxi 214122, Jiangsu, People's Republic of China}

\begin{abstract}
Single image super-resolution (SISR) is the task of inferring a high-resolution image from a single low-resolution image. 
Recent research on super-resolution has achieved great progress due to the development of deep convolutional neural networks in the field of computer vision. 
Existing super-resolution reconstruction methods have high performances in the criterion of Mean Square Error (MSE) but most methods fail to reconstruct an image with shape edges. To solve this problem, the mixed gradient error, which is composed by MSE and a weighted mean gradient error, is proposed in this work and applied to a modified U-net network as the loss function. The modified U-net removes all batch normalization layers and one of the convolution layers in each block. The operation reduces the number of parameters, and therefore accelerates the reconstruction. Compared with the existing image super-resolution algorithms, the proposed reconstruction method has better performance and time consumption. The experiments demonstrate that modified U-net network architecture with mixed gradient loss yields high-level results on three image datasets: SET14, BSD300, ICDAR2003. Code is available online\footnote {The project is coded by PyTorch and is proposed on \url{https://github.com/MnisterLu/simplifiedUnetSR}} 
\end{abstract}


\end{frontmatter}

\section{Introduction}

Significant progress of neural networks for computer vision has been given to the field of Single Image Super-Resolution (SISR). 
SISR aims to reconstruct a super-resolution image $I^{SR}$ from a single low-resolution image $I^{LR}$. 
This task, refered as super-resolution, finds direct applications in numerous areas such as medical image processing~\cite{greenspan2002mri,peled2001superresolution}, HDTV~\cite{goto2014super}, face recognition~\cite{gunturk2003eigenface,wheeler2007multi}, satellite image processing~\cite{thornton2006sub,tatem2001super,tatem2002super} and surveillance~\cite{zhang2010super,lin2005investigation}. 
Furthermore, the super-resolution task can also be expressed as a one-to-many mapping from low-resolution to high-resolution space for each pixel. 
Super-resolution is an inference problem and the main solution to this problem relies on the statistical model in recent researches~\cite{wang2019deep}.

Recently, deep learning based SISR has attracted wide attention. He~\cite{dong2015image} introduced the first deep learning method for single image super-resolution called Super-Resolution Convolutional Neural Network ({\bf SRCNN}). Then, Fast Super-Resolution Convolutional Neural Network ({\bf FSRCNN})~\cite{dong2016accelerating} built a lightweight framework to solve real-time problems of super-resolution. Super-resolution Generative Adversarial Networks ({\bf SRGAN})~\cite{ledig2017photo} was the first Generative Adversarial Networks ({\bf GAN})~\cite{goodfellow2014generative} for super-resolution task and considered the human subjective evaluation of reconstructed images. Enhanced Deep Residual Networks for Single Image Super-Resolution ({\bf EDSR})~\cite{lim2017enhanced} was the winner of the NTIRE2017 Super-Resolution Challenge Competition and achieved an outstanding reconstruction performance. Deep Back-Projection Networks ({\bf DBPN})~\cite{haris2018deep} exploited iterative up-sampling and down-sampling and providing an error feedback mechanism. 

One of the main disadvantages of previous works was that it is difficult to reconstruct clear boundaries and high gradient components. The other was that the architecture of the previous network lacked of skip connection between the convolution layers at the same depth, which caused the information loss. To solve the problem, an improved U-net with a mixed gradient loss is proposed for SISR. The contribution of the new network can be summarized as follows:

\begin{enumerate}

\item The improved {\bf U-net} network architecture is proposed for super-resolution on a single image. The {\bf U-net}, which has been approved effective for image segmentation, is modified for SISR task by removing all batch normalization layers and one convolution layer in each block. The operation reduces the computation cost and meanwhile keeps the reconstruction performance. The input image is up-scaled for the larger size and build a new convolution block on the large scale, which has a skip-connection with the output block on the same scale. The direct up-scaled images avoid the errors caused by redundant calculations.

\item A mixed gradient error (MixGE) is proposed and applied to SISR loss function. As a combination of mean square error (MSE)~\cite{ephraim1984speech} and mean gradient error (MGE), not only pixel error but also gradient error is considered and the improved loss is called MixGE. In the paper, we use the classic gradient calculation method which was proposed by Sobel\cite{kanopoulos1988design}. The method for MGE computation are presented and analyzed.

\item The modified {\bf U-net} for common scenes task super-resolution yields the outstanding performance over existing methods on SET14 and BSD300 dataset. Moreover, the proposed network successfully outperforms other state-of-art methods on texture tasks such as ICDAR2003 dataset.

\end{enumerate}

%-------------------------------------------------------------------------
\section{Related Work}

The goal of super-resolution methods is to recover a high-resolution image from a low-resolution image~\cite{irani2009super}. Recently most popular SISR methods were implemented by deep learning network instead of traditional mathematical models. This section will analyze and summarize most effective existing super-resolution reconstruction methods which realized by deep learning methods.

%-------------------------------------------------------------------------

\subsection{Super-resoution network}

For super resolution, the methods can be divided into video super-reslution reconstruction, such as {\bf VESPCN}~\cite{caballero2017real}, {\bf VSRCNN}~\cite{kappeler2016video}, and single image super-resolution, which includes {\bf SRCNN}~\cite{dong2015image}, {\bf FSRCNN}~\cite{dong2016accelerating}, {\bf VDSR}~\cite{kim2016accurate}, {\bf DRCN}~\cite{kim2016deeply}, {\bf SRGAN}~\cite{ledig2017photo}, {\bf ESPCN}~\cite{shi2016real}, {\bf EDSR}~\cite{lim2017enhanced} and {\bf DBPN}~\cite{haris2018deep}.

To better solve the distortion problem of high-resolution image reconstruction, He~\cite{dong2015image} presented a network architecture named {\bf SRCNN} which was the earliest proposal to solve super-resolution problem through deep learning network. He demonstrated that a convolution network can be used to learn an efficacious mapping from low-resolution to high-resolution in an end-to-end architecture.

Although the {\bf SRCNN} introduced a new way to solve the super-resolution reconstruction problem by deep learning, this method had 2 obvious limitations.  First, the observation field was too narrow that can not get enough corresponding information. Second, the training of {\bf SRCNN} was hard to converge and easy to be overfitted.

To solve these two limitations, {\bf FSRCNN}~\cite{dong2016accelerating} was proposed, which was different from {\bf SRCNN} in 3 aspects. First, {\bf FSRCNN} adopted the original low-resolution image as input instead of the {\bf bicubic}~\cite{de1962bicubic} interpolate image in {\bf SRCNN}. Second, The de-convolution layer was added at the end of the network for upsampling. The third and the most efficient part was to adapt smaller filter kernels and a deeper network into super-resolution task.

The {\bf VDSR}~\cite{kim2016accurate} came up with a much deeper network. To get a large observation field, {\bf VDSR} chose a convolution kernel of 3$\times$3 in the deep network. Otherwise, very deep networks converged too slowly because of the big number of parameters. In the {\bf VDSR} method, residual learning and gradient clipping were chosen to be the solution to the training problem. 

Taking an interpolated image as the input, {\bf DRCN}~\cite{kim2016deeply} was composed of three modules, namely embedding network for feature extraction, inference network for nonlinear feature mapping, and reconstruction network for SR image generation. The Inference network was a recursive network, that is, data looped through the layer multiple times. Expanding this loop was equivalent to multiple concatenated convolution layers using the same set of parameters.

Generative adversarial networks ({\bf GAN})~\cite{goodfellow2014generative}, which was proposed by Goodfellow in 2014, was set up for estimating generative models via adversarial process. First {\bf GAN} proposed for super-resolution task was {\bf SRGAN}~\cite{ledig2017photo}. 
The experimental results in this work showed that the generator network trained on MSE loss function could output SR images with high Peak signal-to-noise ratio (PSNR) but over-smoothed. 
The output images of {\bf SRGAN} had a better visual effect than other methods without adversarial process. 
The work proposed a {\bf GAN} that apply a deep residual network ({\bf ResNet})~\cite{he2016deep} with skip-connection. For improving the visual effect of the super-resolution reconstructed results, they put forward the perceptual loss instead of the MSE loss function.

To solve the high computational complexity of deep network, {\bf ESPCN}~\cite{shi2016real} was proposed for super-resolution task with a much higher processing speed than previous methods. The core concept of {\bf ESPCN} was the sub-pixel convolutional layer. The input to the network was the original low-resolution image. After passing through two convolutional layers, the result feature image was the same size as the input image, and the feature channel was $r^2$ where $r$ was the magnification scale. Feature images of size $r^2\times H\times W$ were re-arranged into high-resolution images of size $1\times rH\times rW$. The operation of convolution layer re-arrangement greatly improved the efficiency of de-convolution and the {\bf ESPCN} had a better performance on super-resolution tasks.

{\bf EDSR}~\cite{lim2017enhanced} was the winner of the NTIRE2017 Super-Resolution Challenge Competition. As stated in the paper, the most significant performance improvement of {\bf EDSR} was to remove the batch normalization layers of {\bf SRResNet}, which can expand the size of the model to improve the quality of the results.

Haris~\cite{haris2018deep} proposed Deep Back-Projection Networks ({\bf DBPN}), applying iterative up-sampling and down-sampling and exploiting an error feedback mechanism for projection errors in every block. This work constructed mutually-connected up-sampling and down-sampling blocks, which represented different parts of low-resolution and high-resolution components.

\subsection{Problem analysis}

\begin{figure*}[th]
\begin{center}
%\fbox{\rule{0pt}{2in} \rule{.9\linewidth}{0pt}}
	\includegraphics[width=1\linewidth]{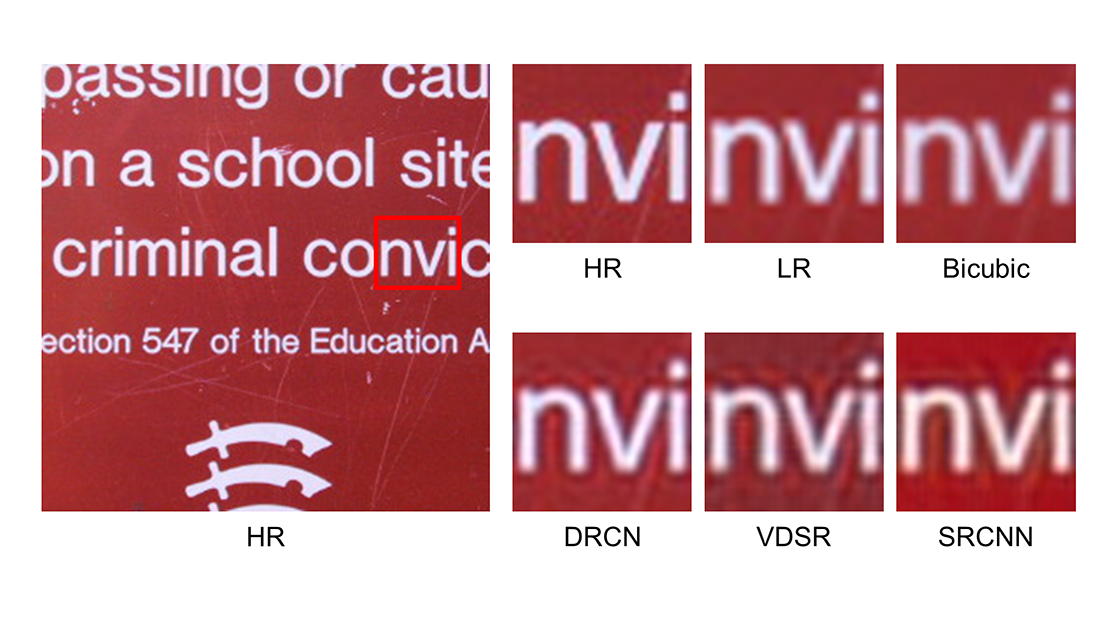}
\end{center}
   \caption{Super-resolution results by previous works on ICDAR2003 dataset($\times$4).}
\label{fig:SRdemo2_5}
\end{figure*}

Previous works on SISR reconstruction task achieved a good performance, but there were still many shortcomings and many parts that lacked reasonable explanations. First, previous studies of SISR had not dealt with the problem of blurred edge, shown as reconstructed results by methods of {\bf SRCNN} and {\bf VDSR} in Figure~\ref{fig:SRdemo2_5}, because these researches were limited to MSE loss function. Though the performance of {\bf SRCNN} on MSE loss function was outstanding, subjective observation of the reconstructed images was blurred. Therefore, gradient components should be considered into the SISR task.

Second, most convolutional layers of SISR deep networks were directly connected which led to a part of data loss in low-dimensional feature layers. High-resolution images which only reconstructed by high-level semantic information lost most basic fine-grained texture details. It is necessary to build a skip-connection between same depth layers to combine high-level semantic information with low-level fine-grained texture details.

Third, according to the impressive analysis of {\bf VDSR}~\cite{kim2016accurate} which concluded that better performance could be achieved with deeper network, the number of parameters of the most SISR network were very large due to very deep networks were used to obtain high performance. The networks suffered from large computation cost which makes it difficult to run in real-time.

%------------------------------------------------------------------------
\begin{figure}[th]
\centering
\subfigure[U-net network architecture.]{
	\includegraphics[width=1\linewidth]{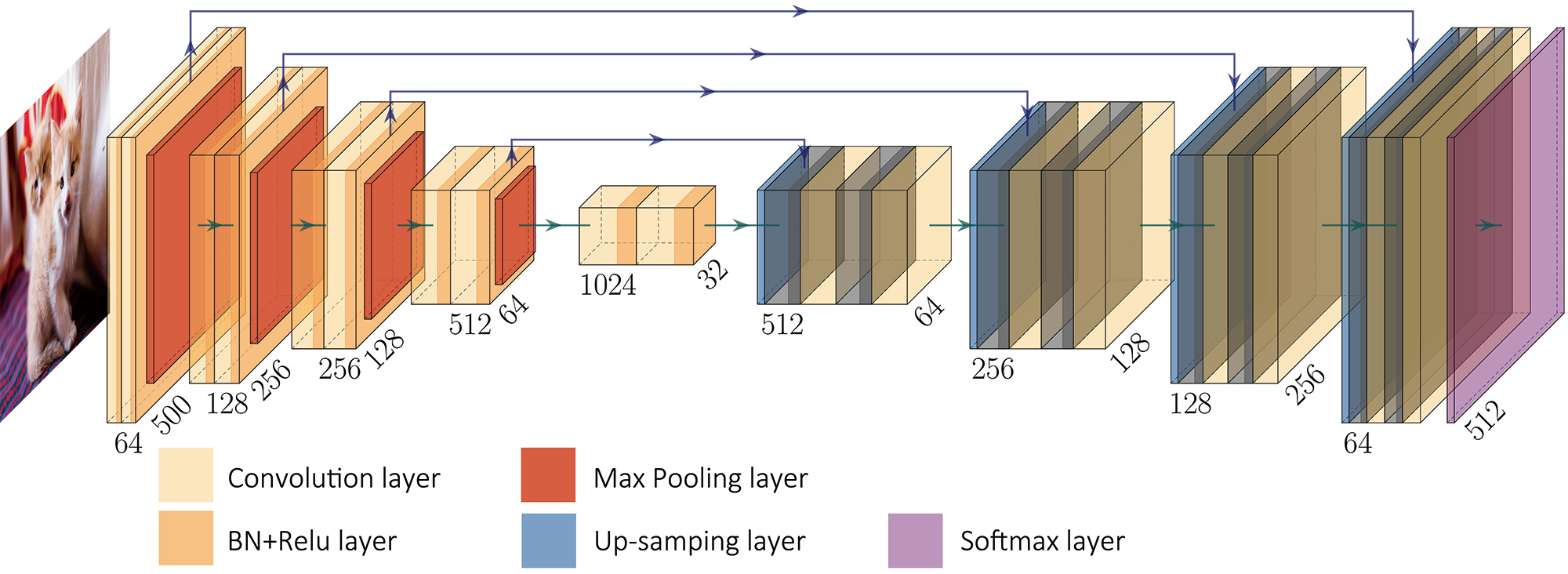}
	\label{fig:Unet}
%\caption{fig1}
}
\quad
\subfigure[modified U-net network architecture.]{
	\includegraphics[width=1\linewidth]{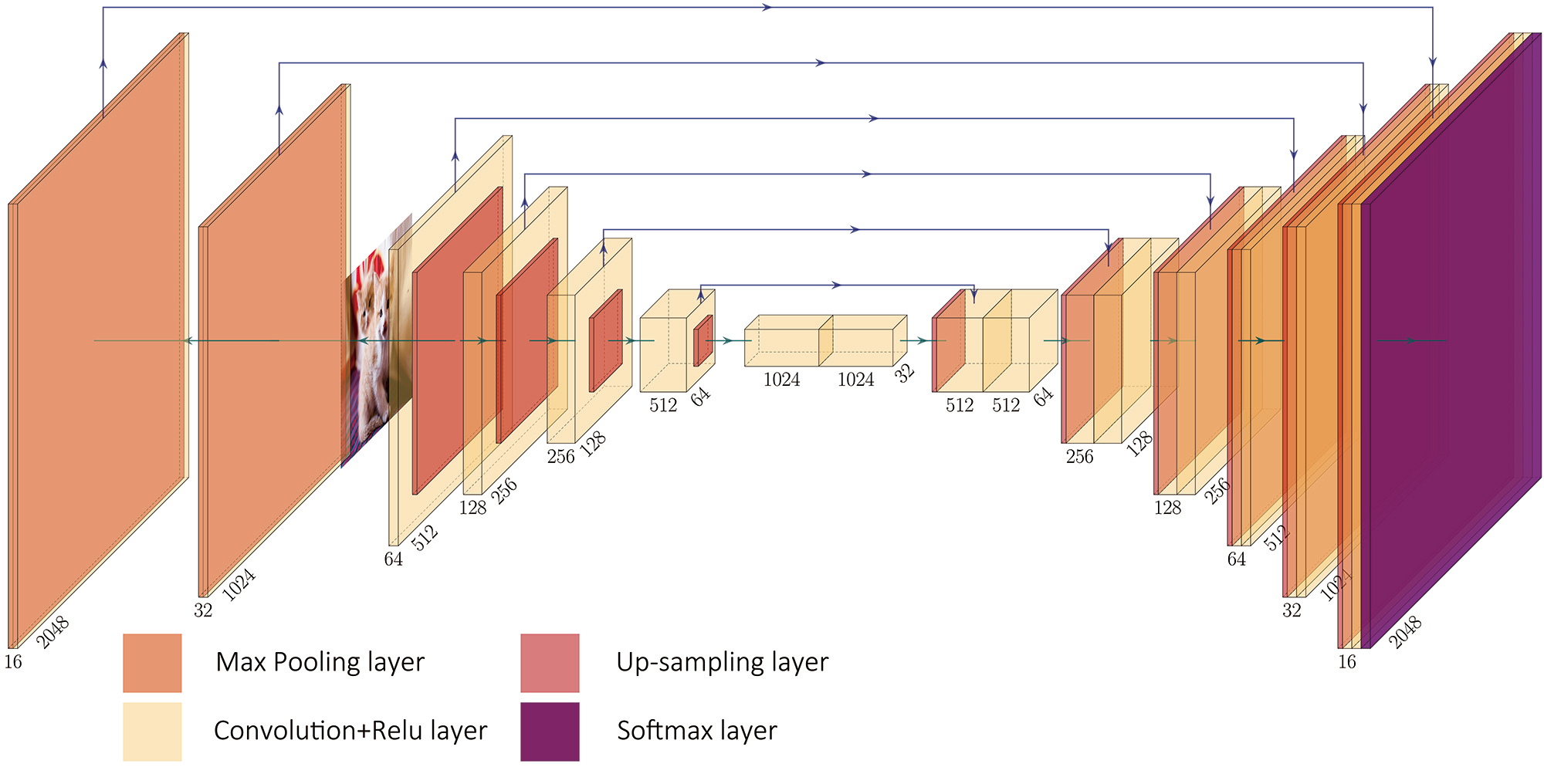}
	\label{fig:SimpleUnet}
}
	\caption{Original U-net and modified U-net network architecture}
\end{figure}

\section{Method}

The task of the single image super-resolution was to reconstruct a super-resolution image $I^{SR}$ from a low-resolution image $I^{LR}$. To solve this single image super-resolution problem, network architecture is constructed where input image $I^{LR}$ is represented as real-valued tensors of size $H\times W\times C$ and output image $I^{SR}$ represented as a tensor of size $rH\times rW\times C$.

This work proposes an improved network called modified U-net and a mixed gradient loss function which combines with Mean Square Error and mean gradient error. 

\subsection{Modified U-net}

{\bf U-net}~\cite{ronneberger2015u} was first proposed by Ronneberger and widely used in the field of semantic segmentation. Ronneberger showed that the {\bf U-net} can be trained as a end-to-end system from extremely few images. Meanwhile, this network architecture surpassed the previous state-of-the-art method on the ISBI challenge for semantic segmentation of neuronal structures.

The {\bf U-net} architecture is illustrated in Figure~\ref{fig:Unet}. The left side is the contracting path which is used for extracting features and the right side is called expansive path for decoding. 
The contracting path contains the continuous part of two 3$\times$3 kernels and each part is followed by a Rectified Linear Unit (ReLU) layer and then a 2$\times$2 max-pooling operation with stride 2 for down-sampling. 
So downscale of each contracting block is 2. 
Each step in the expansive path includes an up-sampling of the feature map, which followed by a 2$\times$2 convolution that halves the number of feature channels, and two 3$\times$3 convolution kernels, each followed by a ReLU layer.

The most important contribution of {\bf U-net} is to put forward the skip-connection. 
The {\bf U-net} network architecture is proposed after the Fully Convolutional Network ({\bf FCN})\cite{long2015fully} and modified in a creative method that it yields the state-of-the-art segmentation in medical image processing at that time. 
Compared with {\bf FCN}, the first main difference is {\bf U-net} is symmetric and the second one is the skip connections between the down-sampling path and the up-sampling path employ a concatenation operation instead of a direct sum. 
Skip connections in the network aim to provide the original local information to the global information when up-sampling. 
Because of the symmetry of {\bf U-net} network architecture, the {\bf U-net} has numerous feature maps in up-sampling blocks that allows the efficacious transmission of information.

As illustrated in Figure~\ref{fig:SimpleUnet}, the modified {\bf U-net} designed in this work for SISR has three differences comparing with the original {\bf U-net}. First, it removes all batch normalization layers and one of the convolution layers in each block. The reason for this is that super-resolution reconstruction is a pixel-level task because the solution to interpolation problem is mainly consider pixels in a certain area. Directly magnified images avoid the errors caused by redundant calculations, which can get closer to real results.

Second, the input image is up-scaled for the larger size and build a new convolution block on the larger scale, which has a skip-connection with the output block on the same scale. The direct up-scaled images avoid the errors caused by redundant calculations, which can get much closer to ground truth. Each addition of upscale layer means expand the size into twice. In other words, the network should have 3 upscale layers if upscale size is 8 and similarly 2 upscale layers for upscale size is 4.

Third, the depth of the original {\bf U-net} is fixed to 4, which means there are 4 downscale blocks and corresponding 4 upscale blocks. Previous work~\cite{ronneberger2015u} has shown that the deeper network leads to the better performance and higher computation cost in most situations. In our work, the depth of the modified {\bf U-net} for SISR is discussed in section 4, showing a trade off between reconstruction accuracy and computation cost.

\subsection{Mixed Gradient Error}

The aim of SISR is to learn a mapping function $f$ for generating an high-resolution image $\hat{Y}=f(X)$ from a low-resolution image $X$ that is close to the ground truth image $Y$. MSE~\cite{allen1971mean} is widely used as most loss function, which measures the average of the squares of each pixel errors in super-resolution reconstruction. MSE can be shows as follows:

\begin{equation}
\begin{split}
MSE=\frac{1}{n}\frac{1}{m}\sum_{i=1}^{n}\sum_{j=1}^{m}(Y(i,j)-\hat{Y}(i,j))^{2}
\end{split}
\end{equation}
\noindent where $n$ is the number of horizontal pixels and $m$ is the number of vertical pixels.

To solve the gradient error measurement problem, we introduce classic gradients to the SISR loss function. In our work, Sobel operator\cite{kanopoulos1988design} is used for gradient calculation, that is,

\begin{equation}
G_{x}=Y*
\begin{bmatrix}
-1 & -2 & -1\\ 
 0 &  0 & 0\\ 
 1 &  2 & 1
\end{bmatrix}
\end{equation}

\noindent where $*$ is the convolution operation.

The gradient map $G$ in $y$ direction of the ground truth image $Y$ shows below:

\begin{equation}
G_{y}=Y*
\begin{bmatrix}
-1 &  0 & 1\\ 
-2 &  0 & 2\\ 
-1 &  0 & 1
\end{bmatrix}
\end{equation}

Then we combine the gradient value of $x$ and $y$ direction as follows:

\begin{equation}
G(i,j)=\sqrt{G_{x}^{2}(i,j)+G_{y}^{2}(i,j)}
\end{equation}

Meanwhile, the gradient map $\hat{G}$ can be calculated in the same way. The aim of measure the gradient error is to learn a sharp edge which is close to the groud truth edge. The Mean Gradient Error (MGE) shows as follows:

\begin{equation}
\begin{split}
MGE=\frac{1}{n}\frac{1}{m}\sum_{i=1}^{n}\sum_{j=1}^{m}(G(i,j)-\hat{G}(i,j))^{2}
\end{split}
\end{equation}

After achieving the MGE, it should be emphasized in the mixed gradient error. As the main component of mixed gradient error, Mean Square Error forms a Mixed Gradient Error (MixGE) by adding Mean Gradient Error with a weight of $\lambda_{G}$.

\begin{equation}
\begin{split}
MixGE(Y,\hat{Y})=MSE+\lambda_{G}MGE
\end{split}
\end{equation}

%-------------------------------------------------------------------------
\section{Experiment}
\subsection{Dataset}

ICDAR2003~\cite{karatzas2013icdar} is a dataset of the ICDAR Robust Reading Competition for text detection and recognition tasks. For the single image super-resolution task, there is not a commonly used dataset. The ICDAR2003 dataset consists of 258 training images and 249 testing images, which contains texts in most of the complex circumstances in common life. Because of the resolution of images varies from 422$\times$102 to 640$\times$480, we resize them into 224$\times$224 with {\bf bicubic} interpolation. This network is also compared with other existing methods on standard benchmark dataset: SET14~\cite{bevilacqua2012low}, BSD300~\cite{martin2001database}.

\subsection{Evaluation method}
Two commonly used performance metric are employed for evaluation and comparsion:  PSNR~\cite{huynh2008scope} and Structural Similarity Index (SSIM)~\cite{hore2010image}.
The super-resolution results are evaluated with the criteria of PSNR~\cite{huynh2008scope} and SSIM~\cite{hore2010image} on three channels in RGB colour space. The criterion of PSNR is based on the error between each corresponding pixels. Because this criterion only consider the numerical error, most high PSNR results do not have good visual performance. Therefore, the criterion of SSIM~\cite{hore2010image} observes the distortion of the image by comparing the changes in the image structure information, thereby obtaining an objective quality evaluation.
The criteria of PSNR and SSIM are all based on luminance. The higher the value of these criteria, the better the performance of image reconstruction.

Two criteria are described as follows. Let $Y$ donate the ground truth and $\hat{Y}$ donate the reconstructed high-resolution images respectively.

\begin{equation}
\begin{split}
MSE=\frac{1}{MN}\sum_{i=1}^{M}\sum_{j=1}^{N} (\hat{Y}(i,j)-Y(i,j))^{2}
\end{split}
\end{equation}

\begin{equation}
\begin{split}
PSNR(Y,\hat{Y})=10log_{10}\frac{255^{2}}{MSE}
\end{split}
\end{equation}

The criterion of SSIM between patches $P_{\hat{Y}}$ and $P_{Y}$ at the same location on ground truth images $\hat{Y}$ and reconstructed high-resolution image $Y$ is defined as

\begin{equation}
\begin{split}
SSIM(P_{\hat{Y}},P_{Y})=\frac{(2\mu_{ P_{\hat{Y}}}\mu_{P_{Y}}+c_{1})(2\sigma_{ P_{\hat{Y}}}\sigma_{  P_{Y}}+c_{2})}{(\mu_{P_{\hat{Y}}}^{2}+\mu_{P_{Y}}^{2}+c_{1})(\sigma_{P_{\hat{Y}}}^{2}+\sigma_{P_{Y}}^{2}+c_{2})}
\end{split}
\end{equation}

\noindent where $\mu_{P_{\hat{Y}}}$ and $\mu_{P_{Y}}$ are the mean of patch $P_{\hat{Y}}$ and $P_{Y}$ respectively. Meanwhile, $\sigma_{P_{\hat{Y}}}$ and $\sigma_{P_{Y}}$ are the deviation of patch $P_{\hat{Y}}$ and $P_{Y}$. $c_{1}$ and $c_{2}$ are small constants. Then, The criterion of $SSIM(\hat{Y} ,Y)$ is the average of patch-based SSIM over the image.

\subsection{Implement Details}
These existed SISR methods are evaluated on 3 common used dataset: SET14, BSD300, ICDAR2003. SET14 and BSD300 consist of natural scenes and ICDAR2003 contain various types of texts in a robust common scene. The ground truth high-resolution images are downscaled by {\bf bicubic} interpolation to generate low-resolution and high-resolution image pairs for training and testing on Table.1. We convert all images into RGB colour space and processing the data into three channels.

\begin{table}[th]
\begin{center}
\caption{Image size of different scales.}
{\begin{tabular}{|l|c|c|l|}
\hline
\multirow{2}{*}{Scale} & \multicolumn{2}{|c|}{Image size} \\
\cline{2-3}
	&LR	&HR\\
\hline
$\times$2	&	112$\times$112 	&224$\times$224\\
$\times$4	&	56$\times$56 	&224$\times$224\\
$\times$8	&	28$\times$28 	&224$\times$224\\
\hline
\end{tabular}}
%\end{center}
%\caption{Image size of different scales}
\label{tab:size}
\end{center}
\end{table}

In the training parameter set, the batch of data is set to $1$. Our model is trained by Adam optimizer~\cite{kingma2014adam} with $\beta_{1}=0.9$, $\beta_{2}=0.999$, $\epsilon=10^{-8}$. The learning rate is set to $10^{-3}$ initially and decreases to half every $25$ epoch. The PyTorch implement our models with one RTX 2080 GPU and code is presented online.

\subsection{Results}

\subsubsection{Network analysis}

Before comparison with other state-of-the-art methods, some parameters of the modified {\bf U-net} and MixGE loss function need to be determined. The first part is the depth of modified {\bf U-net}, which influences the performance of super-resolution results and computational complexity. The second one is the MGE weight for MixGE loss function. 50 randomly selected images from training set are taken as validation data for parameter determination..

\begin{figure}[th]
\begin{center}
   \includegraphics[width=0.8\linewidth]{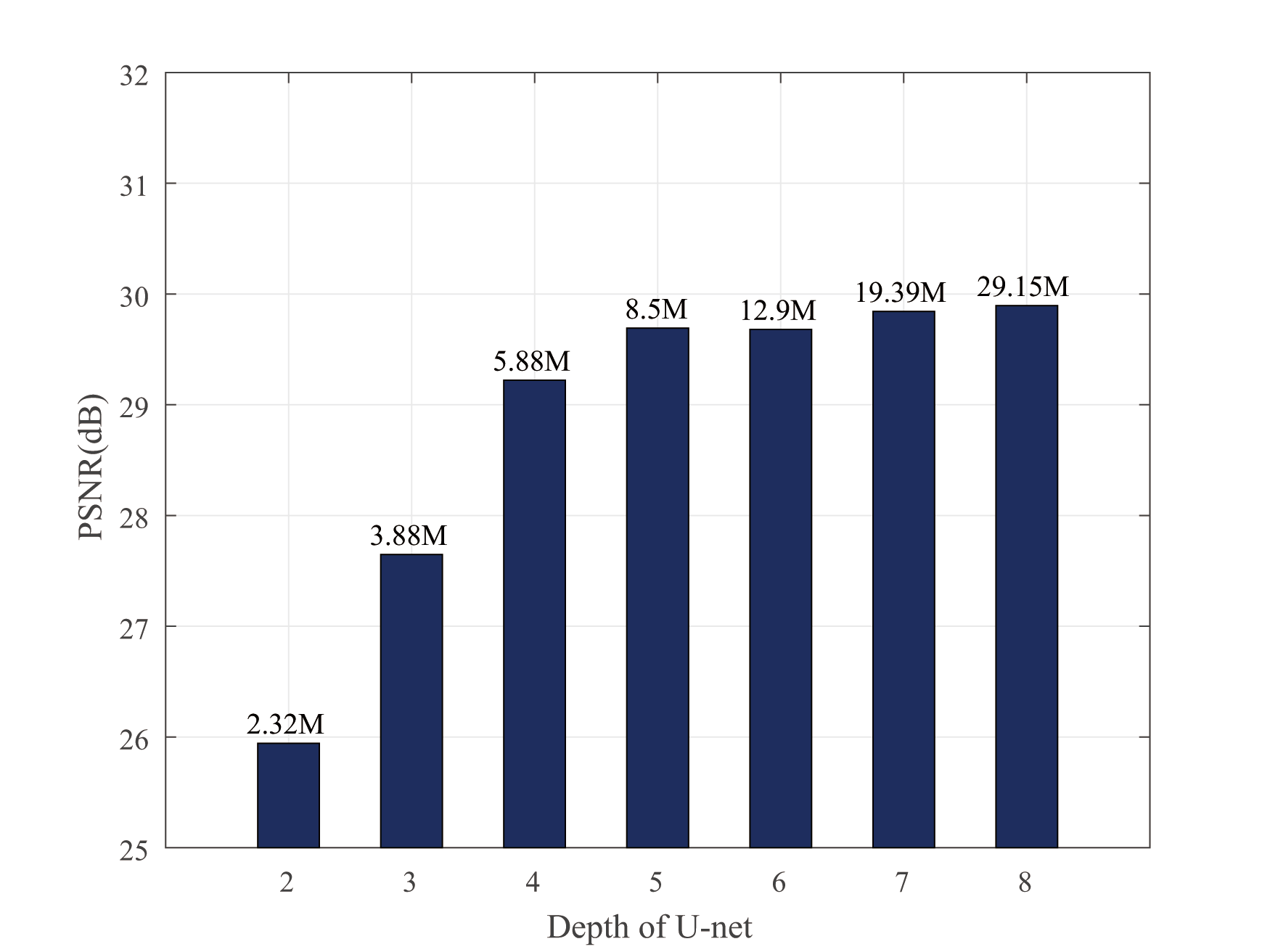}
%\vspace*{8pt}
\end{center}
   \caption{PSNR comparison with different depths of U-net on BSD300($2\times$) dataset.}
\label{fig:depth}
\end{figure}

{\bf Depth analysis.} Figure~\ref{fig:depth} shows the PSNR with increasing depth of U-net, in which the numbers of the networks parameter are shown above each bar. It can be seen from the figure that the performance has a significant improvement from the depth of 2 to 5, while has a slight improvement from the depth of 5 to 8 with large increase in model complexity. Therefore, the depth of 5 is a proper choice that keep a trade-off between reconstruction accuracy and computation cost.

\begin{figure}[th][th]
\begin{center}
   \includegraphics[width=0.8\linewidth]{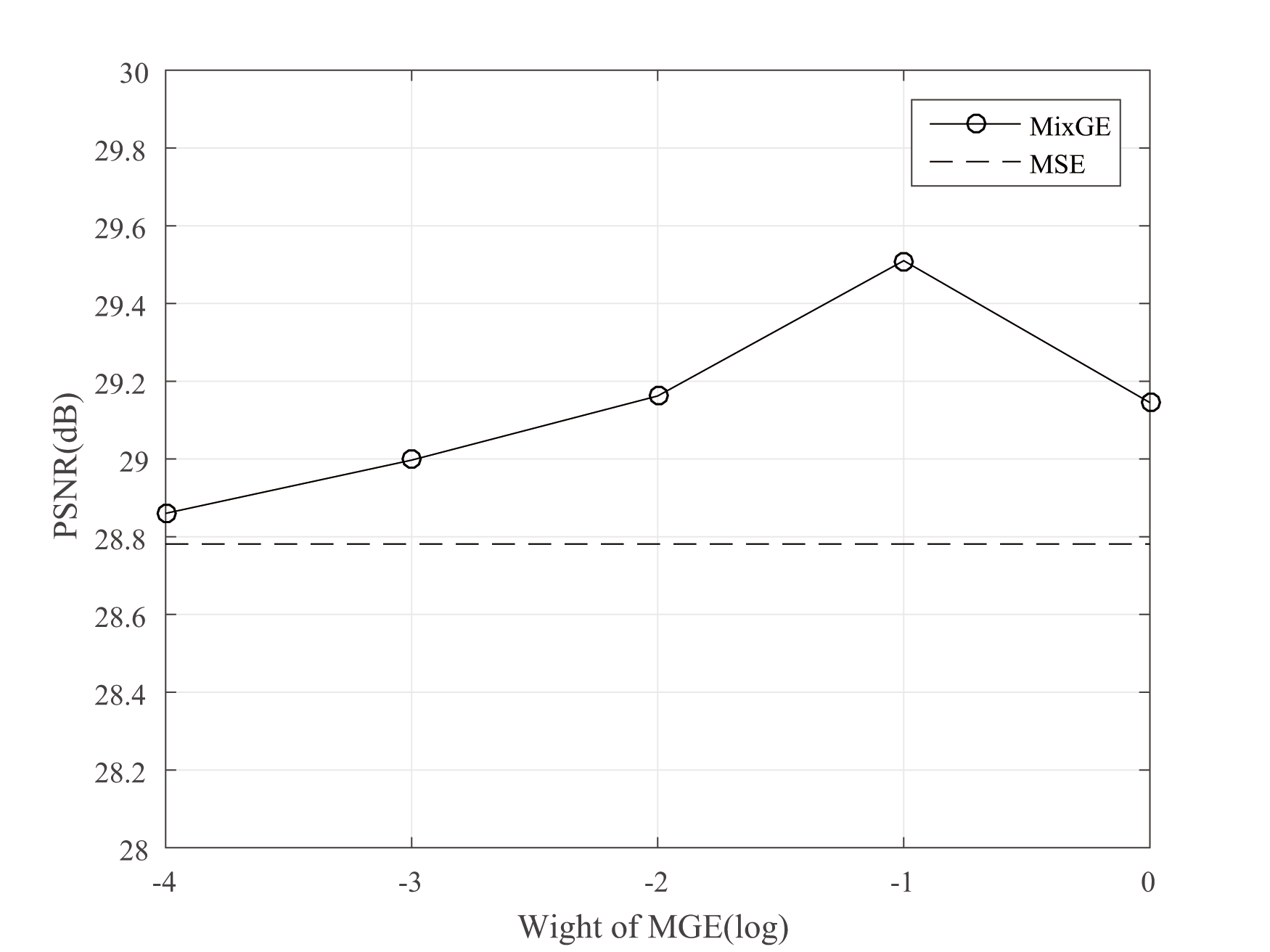}
%\vspace*{8pt}
\end{center}
   \caption{PSNR comparison between MSE loss and MixGE loss with different weights on BSD300($2\times$) dataset.}
\label{fig:Loss}
\end{figure}

{\bf Loss analysis.} For the comparison of MixGE and MSE loss, Figure~\ref{fig:Loss} shows that the results have a great improvement on each dataset if considering the Mean Gradient Error. The method with MixGE loss performs better than that with MSE loss. 

In this experiment, the independent variable, $\lambda_{G}$, was set as $10^{-4}$, $10^{-3}$, $10^{-2}$, $10^{-1}$ and $1$. It can be clearly seen from the Figure~\ref{fig:Loss} that the performance is getting better with the increasing value of $\lambda_{G}$ from $10^{-4}$ to $10^{-1}$ and achieves the peak of performance on $10^{-1}$. This shows that it will greatly improve the performance when choosing the MSE as main component and the MGE as an important auxiliary component. The best performance reaches the PSNR of 29.4dB when the weight equals $0.1$.

It can be seen that MSE loss function is still an irreplaceable loss component in MixGE. MGE becomes an auxiliary component in MixGE to support deep networks to build sharp-edged, gradient-accurate reconstructed images.

\subsubsection{Comparison with the state-of-the-arts}

To assess the performance of our method, we compare the modified {\bf U-net} architecture with existing super-resolution deep learning network, including {\bf SRCNN}~\cite{dong2015image}, {\bf FSRCNN}~\cite{dong2016accelerating}, {\bf VDSR}~\cite{kim2016accurate}, {\bf DRCN}~\cite{kim2016deeply}, {\bf SRGAN}~\cite{ledig2017photo}, {\bf ESPCN}~\cite{shi2016real}, {\bf EDSR}~\cite{lim2017enhanced} and {\bf DBPN}~\cite{haris2018deep}. In this experiment, the performance of {\bf bicubic}~\cite{de1962bicubic} interpolation is chosen to be the baseline method for evaluating performance of deep learning method.

\begin{table*}[!th]
\begin{center}
\caption{Number of parameters comparison between different network architectures.}
%\scalebox{.95}[0.95]
{\begin{tabular}{lcccccccl}
\hline
\multirow{2}{*}{Method}& \multirow{2}{*}{Scale} & \multicolumn{2}{c}{SET14}&\multicolumn{2}{c}{BSD300} &\multicolumn{2}{c}{ICDAR2003}\\
%\cline{3-8}
				&	& 	PSNR 		& 	SSIM		& 	PSNR 		& 	SSIM		& 	PSNR 		& 	SSIM\\
\hline
Bicubic~\cite{de1962bicubic}		& $\times$2	&	24.4523			&	0.8482		&	26.6538		&	0.7924			&	32.9327		&	0.9028	\\
ESPCN	~\cite{shi2016real}			& $\times$2	&	26.7606			&	0.8999		&	28.9832		&	0.8732			&	35.6041		&	0.9243	\\
SRCNN	~\cite{dong2015image}			& $\times$2	&	25.9711			&	0.8681		&	28.6943		&	0.8671			&	35.2711		&	0.9234	\\
VDSR~\cite{kim2016accurate}		& $\times$2	&{\color{red} 28.6617}	&{\color{red} 0.9269}	&	29.3889		&	0.8785			&	36.2323		&	0.9375	\\
EDSR~\cite{lim2017enhanced}		& $\times$2	&	24.0624			&	0.8383		&	28.3119		&	0.8621			&	34.5047		&	0.9258	\\
FSRCNN~\cite{dong2016accelerating}	& $\times$2	&	23.1284			&	0.8123		&	28.7534		&	0.8681			&	35.0533		&	0.9355	\\	
DRCN~\cite{kim2016deeply}			& $\times$2	&	24.4234			&	0.8458		&	27.5089		&	0.8088			&	33.7849		&	0.9185	\\
SRGAN~\cite{ledig2017photo}		& $\times$2	&	23.9553			&	0.8195		&	28.7072		&	0.8633			&	33.2834		&	0.9135	\\
DBPN~\cite{haris2018deep}			& $\times$2	&{\color{blue} 28.4092}	&{\color{blue} 0.9202}&{\color{red} 29.8675}&{\color{red}0.8834}	&{\color{blue} 36.2344}&{\color{blue} 0.9401}\\
UnetSR						& $\times$2	&	26.7241			&	0.8735		&	29.4241		&	0.8813			&	35.7147		&	0.9388	\\
UnetSR+						& $\times$2	&	28.3965			&	0.9198		&{\color{blue} 29.8403}&{\color{blue} 0.8816}	&{\color{red} 37.3673}	&{\color{red} 0.9675}	\\
\hline
Bicubic~\cite{de1962bicubic}	& $\times$ 4	&	19.7167		&	0.6089		&	23.5053		&	0.6157		&	28.1135		&	0.7875	\\
ESPCN~\cite{shi2016real}			& $\times$ 4	&	20.6292		&	0.6333		&	24.4899		&	0.6641		&	29.4861		&	0.8214	\\
SRCNN~\cite{dong2015image}			& $\times$ 4	&	20.5825		&	0.6288		&	24.2232		&	0.6597		&	28.1906		&	0.7661	\\
VDSR~\cite{kim2016accurate}	& $\times$ 4	&	21.4763		&	0.6991		&	24.7077		&	0.6816		&{\color{blue}30.5267}	&{\color{blue}0.8321}	\\
EDSR~\cite{lim2017enhanced}		& $\times$ 4	&	19.9784		&	0.6269		&	23.9192		&	0.6513		&	27.9723		&	0.7101	\\
FSRCNN~\cite{dong2016accelerating}	& $\times$ 4	&	19.3255		&	0.5941		&	24.2499		&	0.6599		&	28.0231		&	0.7652	\\
DRCN~\cite{kim2016deeply}			& $\times$ 4	&	19.7077		&	0.6078		&	23.3462		&	0.6132		&	27.7174		&	0.7764	\\	
SRGAN	~\cite{ledig2017photo}	& $\times$ 4	&	19.3877		&	0.5976		&	24.1675		&	0.6485		&	27.5605		&	0.7654	\\
DBPN~\cite{haris2018deep}			& $\times$ 4	&{\color{red}21.7657}	&{\color{red}0.7171}	&{\color{red}25.0644}	&{\color{red}0.6967}	&	29.8832		&	0.8224	\\
UnetSR	& $\times$ 4	&	20.8891		&	0.6693		&	24.8332		&	0.6843		&	29.3374		&	0.8202	\\
UnetSR+	& $\times$ 4	&{\color{blue}21.6825}	&{\color{blue}0.7112}	&{\color{blue} 24.9522}		&{\color{blue}0.6901}		&{\color{red} 31.8966}	&{\color{red} 0.8898}	\\
\hline
Bicubic~\cite{de1962bicubic}	& $\times$ 8	&	16.1132		&	0.3673		&	21.3115		&	0.4933		&	24.3856		&	0.6831	\\
ESPCN~\cite{shi2016real}		& $\times$ 8	&	16.3441		&	0.3628		&	21.6447		&	0.5064		&	25.1132		&	0.6964	\\
SRCNN~\cite{dong2015image}			& $\times$ 8	&	16.3853		&	0.3614		&	21.8101		&	0.5075		&	22.6281		&	0.6103	\\
VDSR~\cite{kim2016accurate}		& $\times$ 8	&{\color{blue}16.7994}&	0.4095		&	21.9697		&	0.5181		&	25.6303		&	0.7104	\\
EDSR~\cite{lim2017enhanced}		& $\times$ 8	&	15.7257		&	0.3209		&	21.6573		&	0.5067		&	23.5578		&	0.5987	\\
FSRCNN~\cite{dong2016accelerating}	& $\times$ 8	&	14.5788		&	0.2541		&	21.3311		&	0.5011		&	22.5721		&	0.6155	\\
DRCN~\cite{kim2016deeply}			& $\times$ 8	&	16.1497		&	0.3685		&	21.2771		&	0.4934		&	24.2561		&	0.6725	\\
SRGAN~\cite{ledig2017photo}		& $\times$ 8	&	15.7133		&	0.3221		&	21.8766		&	0.5121		&	23.5621		&	0.6425	\\
DBPN~\cite{haris2018deep}			& $\times$ 8	&	16.7398		&{\color{red}0.4122}	&{\color{red}22.0577}	&	0.5229		&{\color{blue}26.3482}&{\color{blue}0.7196}\\
UnetSR						& $\times$ 8	&	16.7001		&	0.4093		&	21.9865		&{\color{blue}0.5231}	&	25.7734		&	0.7106	\\
UnetSR+						& $\times$ 8	&{\color{red}17.8289}	&{\color{blue}0.4103}	&{\color{blue}22.0368}&{\color{red}0.5235}	&{\color{red}28.2512}	&{\color{red}	0.8101}\\
\hline
\end{tabular}}
\end{center}
%\caption{Super-resolution performance of different methods on SET14, BSD300, ICDAR2003 dataset.}
\label{tab:main}
\end{table*}

Table.2 shows the super-resolution performance of different methods on SET14, BSD300, ICDAR2003 dataset, where {\bf UnetSR} denotes the modified {\bf U-net} with MSE and {\bf UnetSR+} denotes the modified {\bf U-net} with MixGE. In the Table.2, the best performance is marked in red and the second best performance is marked in blue in each row. It can seen clearly from the table that {\bf UnetSR+} has the highest PSNR, which means the best performance, at upscale size 2 and 8 on ICDAR2003 dataset. Meanwhile, {\bf UnetSR+} has the second best performance when upscale size is 4.

We average values of PSNR and SSIM of different datasets and upscale sizes for each method. The results are shown in Table 3, together with parameter numbers of each network. 
It is apparent from the Table.3 that the proposed {\bf UnetSR+} achieves the best reconstruction accuracy on average. Commonly, the ISIR performance improves with more network size, while the {\bf UnetSR+} only has $36\%$ parameter numbers of {\bf DBPN}, but the experimental performance exceeds {\bf DBPN} $2.25\%$ on PSNR and $2.47\%$ on SSIM. 
In addition, it can be seen from the table that the proposed {\bf UnetSR+} shows a distinguished trade-off between reconstruction accuracy and model complexity. For example, comparing with {\bf SRGAN}, {\bf DBPN} increased model size by more than 2 times with PSNR increased $8.3\%$ and SSIM increased $9.04\%$, while the {\bf UnetSR+} improves $10.76\%$ on PSNR and $11.73\%$ on SSIM with parameters increased only $30\%$.

\begin{table*}[!th]
\begin{center}
\caption{Average results and number of parameters comparison between different network architectures.}
{\begin{tabular}{lcccl}
\hline
%\cline{3-4}
Method	&	Number of parameters& 	PSNR 		& 	SSIM	\\
		&		(M)			& 	(dB)		& 		\\
\hline
Bicubic	&	 None		&	24.1316 			&	0.6777 		\\
ESPCN		&	 0.08 	&	25.4506 			&	0.7091 		\\
SRCNN		&	 0.17		&	24.8618 			&	0.6880 		\\
VDSR		&	 0.22	&	26.1548 			&	0.7326 		\\
EDSR		&	 0.78	&	24.4100 			&	0.6712 		\\
FSRCNN	&	 0.03	&	24.1128 			&	0.6673 		\\
DRCN		&	 0.11	&	24.2413 			&	0.6783  		\\	
SRGAN		&	 6.54	&	24.2460 			&	0.6761 		\\
DBPN		&	 23.21	&{\color{blue}26.2633} 	&{\color{blue}0.7372}	\\
UnetSR	&	 8.50		&	25.7092 			&	0.7234 		\\
UnetSR+	&	 8.50		&{\color{red}26.8546}		&{\color{red}0.7554}	\\
\hline
\end{tabular}}
\end{center}
\end{table*}

\begin{figure}[th]
\begin{center}
%\fbox{\rule{0pt}{2in} \rule{.9\linewidth}{0pt}}
	\includegraphics[width=1\linewidth]{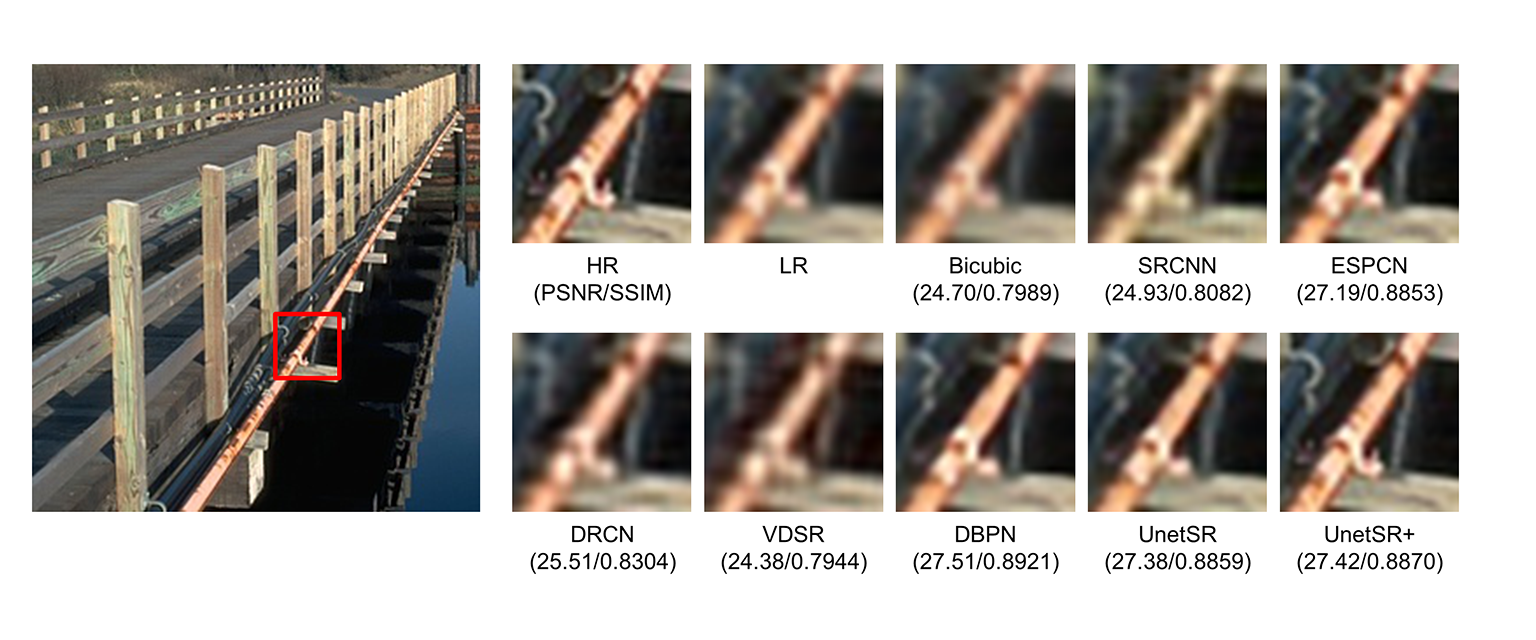}
\end{center}
   \caption{Super-resolution results on BSD300 dataset($\times$2).}
\label{fig:SRdemo2_1}

\end{figure}

\begin{figure}[th]
\begin{center}
%\fbox{\rule{0pt}{2in} \rule{.9\linewidth}{0pt}}
	\includegraphics[width=1\linewidth]{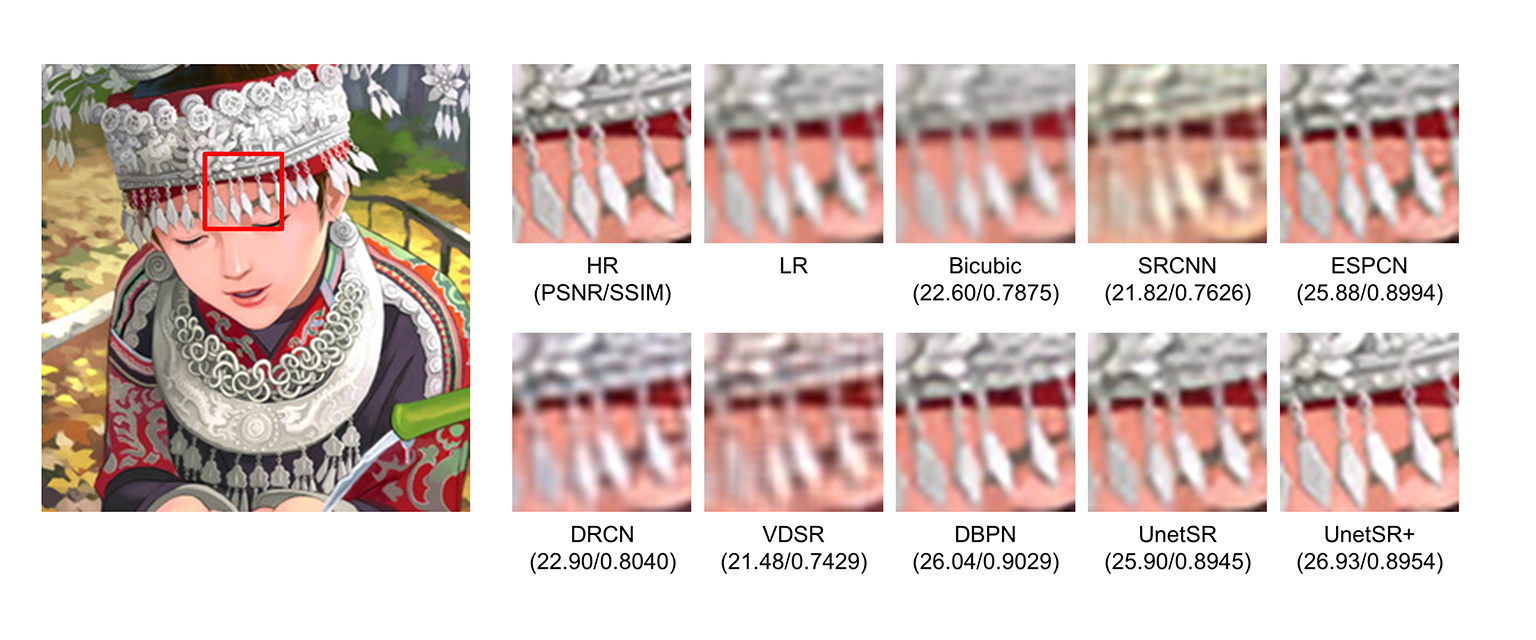}
%\vspace*{8pt}
\end{center}
   \caption{Super-resolution results on SET14 dataset($\times$2).}
\label{fig:SRdemo2_3}
\end{figure}

\begin{figure}[th]
\begin{center}
%\fbox{\rule{0pt}{2in} \rule{.9\linewidth}{0pt}}
	\includegraphics[width=1\linewidth]{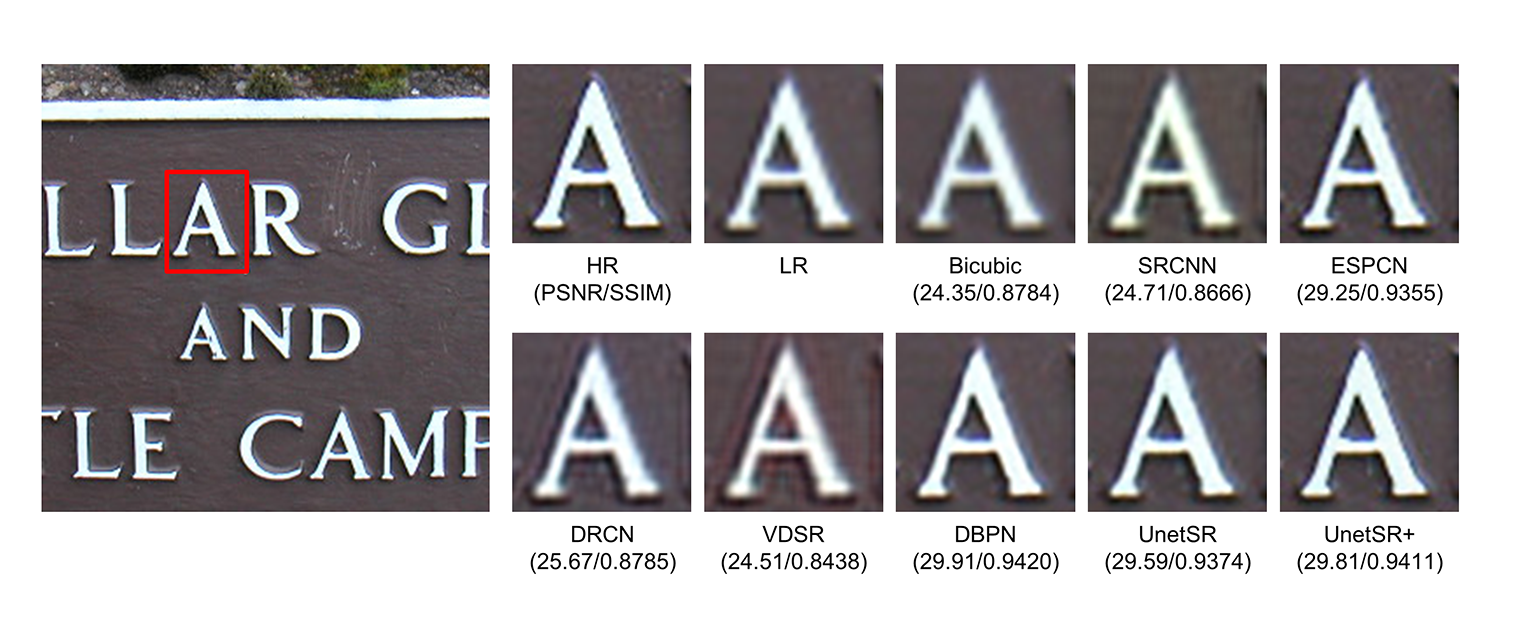}
%\vspace*{8pt}
\end{center}
   \caption{Super-resolution results on ICDAR2003 dataset($\times$4).}
\label{fig:SRdemo2_7}
\end{figure}

\begin{figure}[th]
\begin{center}
%\fbox{\rule{0pt}{2in} \rule{.9\linewidth}{0pt}}
	\includegraphics[width=1\linewidth]{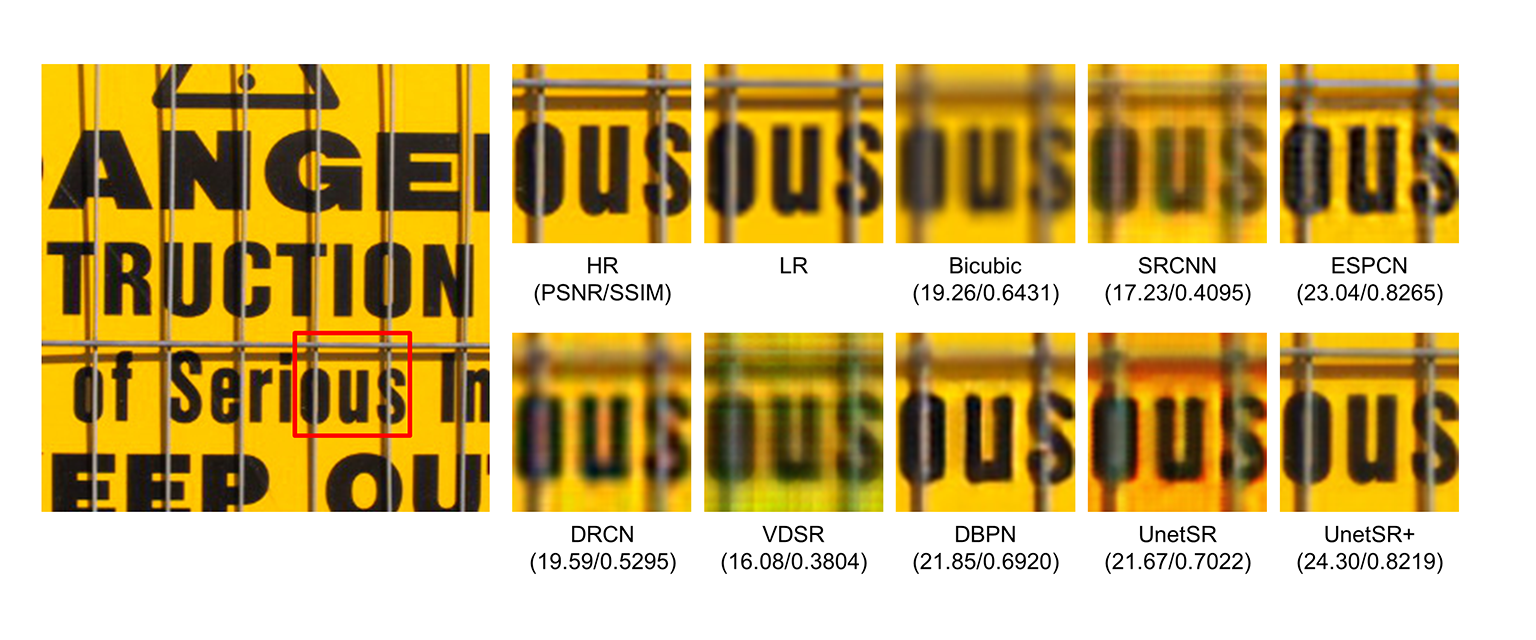}
%\vspace*{8pt}
\end{center}
   \caption{Super-resolution results on ICDAR2003 dataset($\times$8).}
\label{fig:SRdemo4_4}
\end{figure}

For subjective visual evaluation, we compare reconstruction examples with different deep learning methods of SISR for common scenes task on SET14 and BSD300 dataset in Figures~\ref{fig:SRdemo2_1} and ~\ref{fig:SRdemo2_3} and examples for texture task on ICDAR2003 dataset in Figures~\ref{fig:SRdemo2_7} and ~\ref{fig:SRdemo4_4}. 
In Figure~\ref{fig:SRdemo2_1}, super-resolution results on BSD300 dataset for common scenes tasks with the scale of 2 are shown with up-scaled details on the right. 
As illustrated in the figure, the iron pin can be figured out in four methods, which is {\bf ESPCN}, {\bf DBPN}, {\bf UnetSR} and {\bf UnetSR+}. However, the pip strap, a much smaller detail, only reconstructed perfectly by the method of {\bf UnetSR+}. 
It is also clear from the Figure~\ref{fig:SRdemo2_3} that though methods of {\bf ESPCN}, {\bf DBPN}, {\bf UnetSR} and {\bf UnetSR+} reconstruct four distinguishable pendants, only the method of {\bf UnetSR+} provides clear and proper edges for pedants.

For texture task, super-resolution results on ICDAR2003 dataset on the scale of 4 are shown in Figure~\ref{fig:SRdemo2_7}. The left original image has clear convex white characters on a uniform dark brown background. Under this classic super-resolution scene, the reconstructed images of {\bf Bicubic}, {\bf SRCNN}, {\bf ESPCN}, {\bf DRCN} and {\bf VDSR} produced a serious overlap phenomenon. Contrary to these results, each image reconstructed by methods of {\bf DBPN}, {\bf UnetSR} and {\bf UnetSR+} produced a visible border. The method of {\bf UnetSR+} not only restores the sharp border, but also even restores the convexity of the original character. In Figure~\ref{fig:SRdemo4_4}, it shows super-resolution results on ICDAR2003 dataset with the scale of 8 under a more complex scene. The methods of {\bf ESPCN}, {\bf DBPN} and {\bf UnetSR+} are still able to provide distinguishable reconstruction results, in which the {\bf UnetSR+} restores almost all the details and boundaries with sharper edges.

%
%\begin{figure}[bh]
%\begin{center}
%%\fbox{\rule{0pt}{2in} \rule{0.9\linewidth}{0pt}}
%   \includegraphics[width=1\linewidth]{param.eps}
%\vspace*{8pt}
%\end{center}
%	\caption{Number of parameters comparison between different network architectures.}
%\label{fig:params}
%\end{figure}
%

%------------------------------------------------------------------------
\section{Conclusion}

{\bf U-net} for image segmentation is modified in the paper and applied to SISR. First, we remove the superfluous part which includes batch normalization layers and one of the convolution layers. Second, the input image is up-scaled for the larger size and the direct up-scaled images avoid the errors caused by redundant calculations. The modified {\bf U-net} network achieve a high value of PNSR with the trade-off the depth of 5. Furthermore, a Mixed Gradient Loss, which is combined with Mean Square Error and Mean Gradient Error, is proposed for sharp edge reconstruction. Experiments show that the proposed network successfully outperforms other state-of-art methods on SET14, BSD300 and ICDAR2003 datasets. In view of the outstanding performance of the work on texture reconstruction task, it is readily applicable to low-resolution texture detection and recognition in future work.

%%%

\section*{Acknowledgments}

This work was supported by the National Natural Science Foundation of China (61573168).

\bibliographystyle{ieeetr}
\bibliography{egbib}



%% References
%% Usage of\cite is as follows:
%% \cite{key}     ==>> [#]
%% \cite[chap. 2]{key} ==>> [#, chap. 2]
%% \cite{key}     ==>> Author [#]




%%  \bibitem{key}...
%%


% \end{thebibliography}

\end{document}